\documentclass[10pt,journal,cspaper,compsoc]{IEEEtran}

\usepackage{amsmath,amssymb}
\usepackage[pdftex]{graphicx}
\usepackage{balance}
\usepackage{booktabs}
\PassOptionsToPackage{hyphens}{url}\usepackage{hyperref} 
\usepackage{algorithm} 
\usepackage{algorithmic}
\usepackage{multirow}
\usepackage{cite}
\usepackage{url}
\usepackage{color}
\usepackage{tabularx}
\usepackage{xspace}
\usepackage{subfig}

\newcommand{\ie}{\emph{i.e.}\xspace}
\newcommand{\eg}{\emph{e.g.}\xspace}

\newcommand{\etc}{\emph{etc.}\xspace}

\newcommand{\vct}[1]{\ensuremath{\boldsymbol{#1}}} 

\newcommand{\set}[1]{\ensuremath{\mathcal{#1}}}
\newcommand{\con}[1]{\ensuremath{\mathsf{#1}}}

\begin{document}

\title{Digital Investigation of PDF Files:\\Unveiling Traces of Embedded Malware}

\author{Davide~Maiorca,~\IEEEmembership{Member,~IEEE,}
	Battista~Biggio,~\IEEEmembership{Senior Member,~IEEE,}
\IEEEcompsocitemizethanks{\IEEEcompsocthanksitem {Preprint of the work accepted for publication in the IEEE Security \& Privacy magazine, Special Issue on Digital Forensics, Nov. - Dec. 2017, \url{http://ieeexplore.ieee.org/stamp/stamp.jsp?arnumber=7854112}}}
\IEEEcompsocitemizethanks{\IEEEcompsocthanksitem The authors are with the Department of Electrical and Electronic Engineering, University of Cagliari, Piazza d'Armi, 09123 Cagliari, Italy.
\IEEEcompsocthanksitem Davide Maiorca: e-mail davide.maiorca@diee.unica.it
\IEEEcompsocthanksitem Battista Biggio: e-mail battista.biggio@diee.unica.it}}%

\IEEEcompsoctitleabstractindextext
{\begin{abstract}
Over the last decade, malicious software (or malware, for short) has shown an increasing sophistication and proliferation, fueled by a flourishing underground economy, in response to the increasing complexity of modern defense mechanisms. PDF documents are among the major vectors used to convey malware, thanks to the flexibility of their structure and the ability of embedding different kinds of content, ranging from images to JavaScript code. Despite the numerous efforts made by the research and industrial communities, PDF malware is still one of the major threats on the cyber-security landscape. In this paper, we provide an overview of the current attack techniques used to convey PDF malware, and discuss state-of-the-art PDF malware analysis tools that provide valuable support to digital forensic investigations. We finally discuss limitations and open issues of the current defense mechanisms, and sketch some interesting future research directions.
\end{abstract}} 

\maketitle

\IEEEdisplaynotcompsoctitleabstractindextext

\section{Introduction}
\label{sect:introduction}

In recent years, the number of services available on the Internet, along with the number of interconnected users, has rapidly increased. This has revolutionized the way society is organized, facilitating the way we communicate, work, and perform our daily activities. 
However, this rapid expansion of the Internet has also exhibited severe drawbacks.
The first is related to the fact that we are essentially dipped into a liquid state in which a vast amount of our personal data -- a \emph{valuable} asset both for companies and for cybercriminals -- is provided in a seamless manner to third-party services, without guarantees on how it will be managed and stored. Second, the proliferation of web services has also drastically increased the number of vulnerable applications exploitable by cybercriminals. 
Cybercrime has become a very profitable activity, and cybercriminals re-invest profits made on the black markets or other illicit activities (\eg, violated online bank accounts) to improve their illegal business. %
The fact that attackers are economically motivated and constantly aim to mislead current cybersecurity systems, is the main reason behind the constant evolution, sophistication and variability of malware and other scams perpetuated over the Internet.

Portable Document Format (PDF) documents have been among the major vectors used to convey malware, thanks to the flexibility of their structure and the ability of embedding different kinds of content, ranging from JavaScript to ActionScript code.
Although Microsoft Office macro-based attacks are now playing a major role in the diffusion of malware, critical vulnerabilities are still being publicly disclosed for Adobe Reader (see, e.g., \texttt{CVE-2017-3010}, \texttt{CVE-2016-1009}). PDF malware thus remains a potential, serious threat for Internet users, as also witnessed by recent research work~\cite{carmony16,Xu16}.

Malware embedding in PDF files can be largely automatized with state-of-the-art tools like Metasploit. PDF files also support embedding of obfuscated or encrypted content, which can be leveraged by an attacker to increase the probability of evading anti-malware mechanisms.
Another reason behind the proliferation of malicious PDF files is that, normally, unexperienced users receiving such files (e.g., as attachments to scam emails) tend to trust and execute them, as they are not commonly known as potential malware vectors.

Due to the inherent flexibility and complexity of the format, and of the variability of the attacks, effectively analyzing and recognizing malicious PDF files has become a compelling challenge, especially from the viewpoint of a forensic analyst.
For these reasons, machine learning has been exploited as a key component in the development of more recent PDF malware detection systems, either to prevent infection of a targeted machine, or to help the analyst during a forensic investigation (after the incident)~\cite{Maiorca,MaiorcaICISSP,Srndic16,Smutz,DBLP:conf/ndss/SmutzS16,Cova,Laskov,corona14-aisec}.
Nevertheless, as machine learning has not been originally designed to operate against intelligent attackers, it is also known that it exposes specific vulnerabilities that can be exploited to evade detection.

In this paper, we first provide an overview of the PDF file format and of the current attacks used to convey PDF malware, through concrete attack examples collected in the wild.
We describe how to perform a forensic analysis of a PDF file to find evidence of embedded malware, using some state-of-the-art software tools.
We then discuss some of the most recent PDF malware detection tools based on machine learning, which can be used to support digital forensic analyses, identifying suspicious files before digging deep into a more detailed manual investigation. 
We discuss their limitations and related open issues, especially in terms of the exploitation of their  vulnerabilities to potentially mislead subsequent forensic analyses.
We finally suggest guidelines for improving the performance of such systems under attack, and sketch promising research directions.

\section{PDF File Format} \label{sect:pdf-file-format}

\begin{figure}[t]
  \centering
  \includegraphics[width=0.48\textwidth]{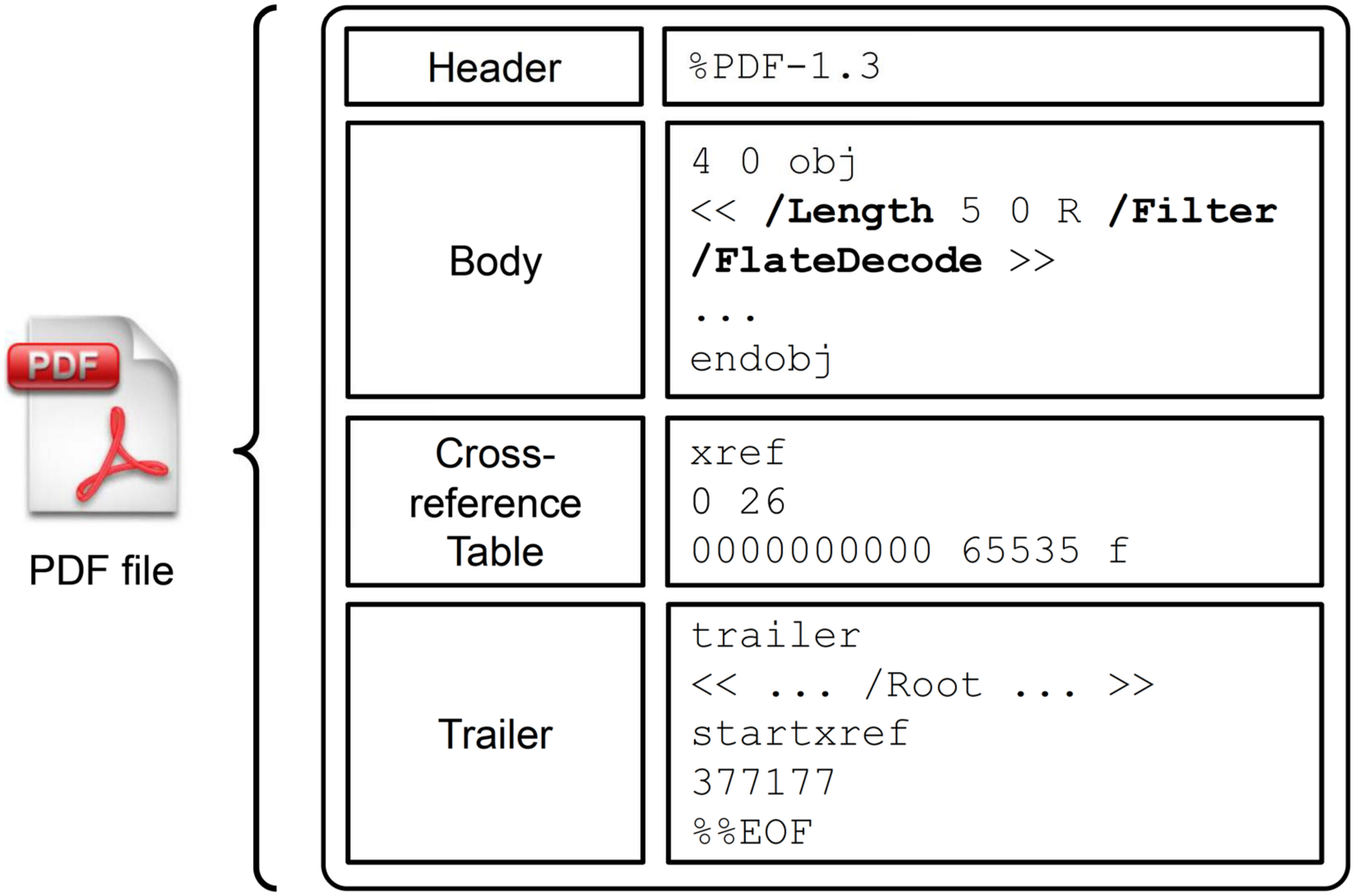}
  \caption{PDF file structure, with examples of header, body, cross-reference table and trailer contents. Object names (\ie, keywords) are highlighted in bold.}
  \label{sect:pdf-file-formatfig:pdf-file}
\end{figure}

PDF is one of the most used format to render documents. Due to the support for third-party technologies such as JavaScript and ActionScript, PDF is widely used not only for visualizing text but also for rendering images, compiling forms, and showing animations.
The typical structure of a PDF file is depicted in Figure~\ref{sect:pdf-file-formatfig:pdf-file}. It consists of four parts: ($i$) the \emph{header}, containing information about the PDF file version; ($ii$) the \emph{body}, containing a number of objects that define the operations performed by the file and the embedded data (\eg, text, images, scripting code); ($iii$) the \emph{cross-reference (x-ref) table}, listing the offset of each object inside the file to be rendered by the reader; 
and ($iv$) the \emph{trailer}, namely, a special object that describes, among others, the first object to be rendered by the reader, identified by the name object \emph{/Root}.

Technically, A PDF file can be seen as a graph of objects that instructs the reader about the operations it has to do to visualize the file content to the user. The PDF file format supports eight types of objects: 
\begin{itemize}
\item
\emph{boolean}, \ie, a variable which can be \texttt{True} or \texttt{False};
\item
\emph{numeric}, \ie, a real or integer value;
\item
\emph{string}, \ie, a sequence of characters between parentheses ( ), or a sequence of hexadecimal characters between angle brackets $<$ $>$;
\item
\emph{name}, \ie, a literal sequence of characters that starts with a forward slash /;
\item
\emph{array}, \ie, a sequence of objects between square brackets [ ];
\item
\emph{dictionary}, \ie, an object composed by a sequence of key-value pairs, enclosed by double angle brackets $<<$ $>>$ (\eg, the \emph{trailer} object is a dictionary); \item
\emph{stream}, \ie, a special object consisting of a dictionary and a sequence of data (typically, compressed text or images), introduced by the keyword \emph{stream}.
\end{itemize}

The aforementioned objects are divided into two categories. Objects that are marked by a number (introduced by the string \emph{objNum 0 obj}) are called \emph{indirect}, whereas objects that are not marked by a number are called \emph{direct}. Indirect objects are typically dictionaries, and can be referenced by other objects with the string \emph{objNum 0 Ref}. An example of indirect object introduced by the string \emph{4 0 obj} is shown in Figure \ref{sect:pdf-file-formatfig:pdf-file}. In this case, the keyword \texttt{/Length} introduces the \emph{size} of the object, whose value is contained in object $5$ (this reference is defined by \emph{5 0 R}). The remaining two keywords define the characteristic of the object, which in this case contains information about the filter used for data compression (\texttt{/FlateDecode}).  The PDF graph mainly contains indirect objects. 

When a reader renders a PDF file, it starts from the trailer object and parses each indirect object (referenced by the \emph{x-ref table}), decompressing its data. In this way, all pages, text, images and scripting code are progressively shown.

\section{PDF Malware}
\label{sect:attacks}

The capability of embedding different kinds of content does not only make the PDF file format a convenient way of legitimately sharing information.
It also gives attackers the possibility of exploiting a larger number of potential vulnerabilities. In fact, PDF malware is multifaceted, conceived to exploit the flexible nature of the PDF file format. 
Typically, JavaScript code, encoded streams and embedded objects (\eg, images, ActionScript code) are used to exploit a vulnerability of the PDF reader and subsequently allow execution of remote code.
In the following, we briefly discuss some popular examples of attacks in which an embedded object (respectively, an image, an executable and a ShockWave Flash file) is used to exploit a vulnerability of the PDF reader. 

The first example exploits the so-called Adobe Reader BMP/RLE heap corruption vulnerability (CVE-2013-2729) to download and install malware from a remote website.\footnote{\url{http://eternal-todo.com/blog/cve-2013-2729-exploit-zeusp2p-gameover}}
In this case, the malicious PDF file contains a form with an encoded bitmap image. When the PDF file is opened, the image is automatically decoded, causing a heap overflow that allows execution of remote code.

Another example, reported by Contagio in $2010$, shows how to execute binary code by simply opening a PDF file (CVE-2010-1240).\footnote{\url{http://contagiodump.blogspot.it/2010/08/cve-2010-1240-with-zeus-trojan.html}}
A code excerpt of the embedded object that implements this attack is reported below.
\begin{verbatim}
	155 0 obj
	<<
	/Type /Action /S /Launch /Win
	<<
	/F (cmd.exe)
	/P (/c echo Dim BinaryStream > vbs1.vbs 
	&& echo Set BinaryStream = 
	CreateObject("ADODB.Stream") >> 
	 ... >>
	endobj
\end{verbatim}
This object executes the (Windows) command prompt (\texttt{cmd.exe}) and uses it to run a Visual Basic script that retrieves and executes the Zeus trojan. 
Notably, execution of binary files has been inhibited in subsequent versions of Adobe Reader, to limit this kind of exploitation. Despite this, and even if this attack is almost seven years old, only $33$ out of the $55$ anti-malware systems used in \texttt{VirusTotal} correctly detect this file as malicious.

Malicious ShockWave Flash (SWF) files and ActionScript code can also be embedded in PDF files to exploit vulnerabilities of the Flash interpreter used by Adobe Reader.
An example is the zero-day vulnerability discovered in 2010 (CVE-2010-1297),\footnote{\url{https://blog.zynamics.com/2010/06/09/analyzing-the-currently-exploited-0-day-for-adobe-reader-and-adobe-flash/}} which was exploited through the execution of a  malicious ActionScript code fragment contained in an embedded SWF file. After exploitation, the infection of the victim machine was completed by running an executable file, also stored as an encrypted stream in the PDF file, which eventually dropped other malware from malicious websites.

Besides embedding external objects, using malicious JavaScript code constitutes the prominent way to attack Adobe Reader. In particular, the goal of the exploit is typically to bypass memory protections such as Data Execution Prevention (DEP) and Address Space Layout Randomization (ASLR) by resorting to a combination of Heap Spraying and Return Oriented Programming (ROP) gadgets. The main idea here is that the attacker fills the Heap with multiple replicas of NOP sleds and shellcode (typically built through ROP gadgets - instructions belonging to existing, legitimate libraries that can be combined to build malicious routines). This is done to increase the probability that, after memory corruption, the execution of the process is redirected to the malicious code. An example of this exploit procedure is the \texttt{CVE-2014-0496} vulnerability, whose full description is reported in \cite{corona14-aisec}.

The aforementioned examples of attack only constitute a small excerpt of the set of possibilities that an attacker has in order to exploit the vulnerabilities of PDF readers.
Nonetheless, they clearly show how sophisticated and different such attacks can be, also highlighting the complexity of the detection task.

\section{Forensic Analysis of PDF Malware}
\label{sect:manual}

From a forensic perspective, assuming that the infection started from a PDF file,
it is essential to depict a basic roadmap that the analyst can follow to identify the suspicious PDF files.
They can be detected (also after infection) by the machine-learning approaches described in the remainder of this manuscript, or identified through some other source of information (e.g., by discussing with the victim the possibility of being phished by a scam e-mail).
Then, their content can be analyzed to identify the actions performed by the malicious code.
Accordingly, the analyst is required to first find the \emph{suspicious indirect objects} (i.e., the malicious scripting code or files embedded in the PDF document) that are responsible for the malware infection. To this end, he/she might employ three different approaches, detailed below. 

\subsection{Keyword-based Analysis}
\label{sect:manual:subsect:key}
The goal here is to extract the content (keywords) of indirect objects to identify the actions performed by the file; \eg, if the keyword \texttt{/JavaScript} is present, the file contains some scripting code. Such analysis does not typically decompress the streams related to the object, but can give the analyst a quick overview of which parts of the file to analyze more in detail. Normally, if no suspicious keyword is present in the file, then this can be considered as safe. 

\texttt{PDFiD}\footnote{\url{https://blog.didierstevens.com/2009/03/31/pdfid/}} to extract \emph{name objects} is a forensic tool for PDF files that can greatly aid this approach. It basically performs textual analysis of all the dictionaries included in the file,  such as objects can be easily visualized with a simple text editor (e.g., \texttt{Notepad}). The result is a list of keyword objects, along with their occurrence in the file. However, such tool can be easily deceived.  First, there is no control on how the objects are connected to each other. This means that the tool can report objects that are never parsed by the reader. Moreover, the tool does not consider the global structure of the PDF file. Hence, it can extract objects from positions in which they could never be parsed by the reader (as it happens in \cite{Srndic14}). 
For these reasons, the results provided by the \texttt{PDFiD} should be further confirmed by other tools/approaches.
 
\subsection{Tree-based Analysis}
\label{sect:manual:subsect:tree}
Here the goal is to reconstruct the PDF file \emph{tree}, i.e., the interconnections among its objects. \texttt{PeePDF}\footnote{\url{http://eternal-todo.com/tools/peepdf-pdf-analysis-tool}} is a publicly-available software that performs this operation automatically. In particular, its analysis is performed as follows: the system first looks for the trailer object (containing the \texttt{/Root} keyword), which is always the first object of the hierarchy. Then, it uses the reference contained next to the \texttt{/Root} keyword to locate the \texttt{/Catalog} object, which is the main object outside the trailer. Each of the subsequent references is then used to reconstruct the tree. Most malicious files are based on trees that finish with objects containing suspicious actions. \texttt{PeePDF} automatically underlines them,  and allow dumping their stream for content analysis. 

\texttt{Origami}\footnote{\url{http://esec-lab.sogeti.com/pages/origami.html}} is very similar to \texttt{PeePDF}, as it also allows to visualize the PDF file structure. It additionally provides routines for encrypting and decrypting files, extracting metadata, \etc


\subsection{Code-based Analysis} 
The goal of the analyst here is to analyze embedded scripting code without focusing on the internals of the PDF file. This analysis is usually performed to unveil the presence of scripting lines related to known vulnerabilities, which can provide clear hints on the maliciousness of the file.  
\texttt{PeePDF} and \texttt{Origami} both have functionalities to automatically detect suspicious strings inside JavaScript codes. However, \texttt{PhoneyPDF} is probably the best software to perform such analysis. In fact, this software (written in Python) first detects objects bearing JavaScript-related keywords. Then, it instruments and executes the extracted JavaScript code with a JavaScript interpreter to point out suspicious functions.
Such analysis is limited by the fact that it is only related to the detection of JavaScript, and ignores other attack possibilities, like SWF file embedding.

\section{Learning-based PDF Malware Detection}
\label{sect:detection}

\begin{figure*}[ht]
  \centering
  \includegraphics[width=0.65\textwidth]{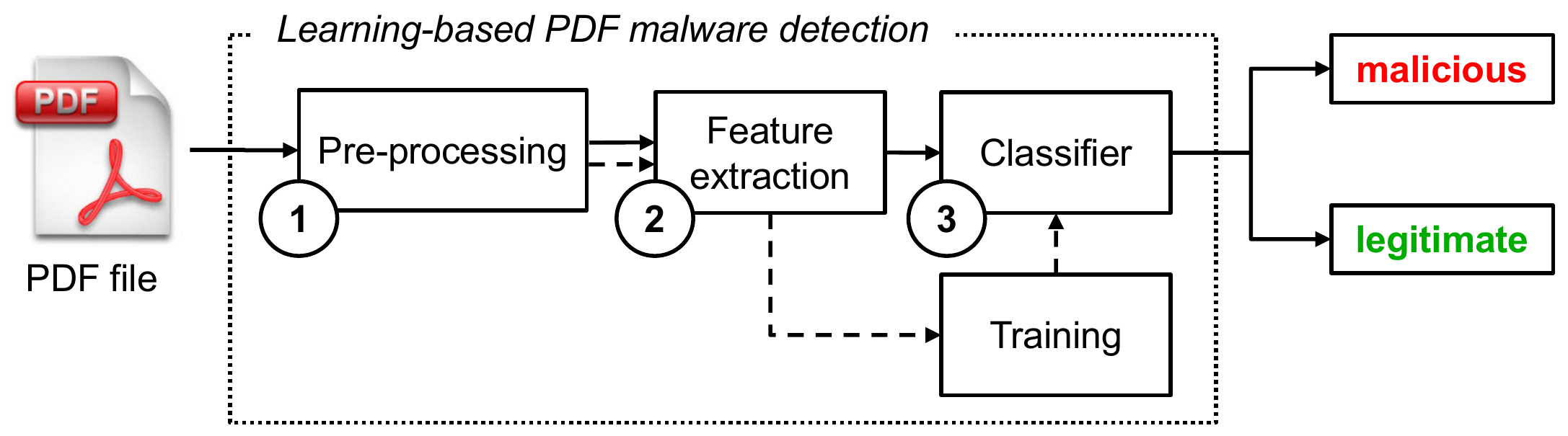}
  \caption{Graphical architecture of a learning-based PDF malware detection tool.}
  \label{fig:system-outline}
\end{figure*}

The aforementioned forensic techniques can be used after the identification of a set of suspicious PDF files, to identify the malware code responsible for the infection and characterize its behavior.
The learning-based PDF malware detection tools discussed in this section have been normally proposed to prevent novel infections, but they can also be used in a forensic investigation, to identify the suspicious PDF files which demand for a subsequent detailed analysis.
Notably, machine learning has been increasingly applied as a key component in recent PDF malware detectors to counter the growing variability and sophistication exhibited by current PDF malware. The design of such tools is based on the three main steps shown in Figure~\ref{fig:system-outline}, and described below.

\begin{table*}[t]
\centering
\caption{An overview of the main characteristics of current PDF malware detectors.} 
\label{tab:pdfDetectors}
\begin{tabular}{@{}lcccc@{}}
\toprule
\multicolumn{1}{l}{\emph{Detector}} & \multicolumn{2}{c}{Pre-processing} & Features & Classifier \\ \midrule
\multicolumn{1}{l|}{\textbf{Wepawet} \cite{Cova}} & \multicolumn{1}{c|}{Dynamic} & JSand & JS-based & Bayesian \\
\multicolumn{1}{l|}{\textbf{PJScan} \cite{Laskov}} & \multicolumn{1}{c|}{Static} & Poppler & JS-based & SVM \\
\multicolumn{1}{l|}{\textbf{Hidost} \cite{Srndic16}} & \multicolumn{1}{c|}{Static} & Poppler & Structural & Random Forest \\
\multicolumn{1}{l|}{\textbf{Lux0R} \cite{corona14-aisec}} & \multicolumn{1}{c|}{Static} & PhoneyPDF & JS-based & Random Forest \\
\multicolumn{1}{l|}{\textbf{Slayer} \cite{Maiorca}} & \multicolumn{1}{c|}{Static} & PDFID & Structural & Random Forest \\
\multicolumn{1}{l|}{\textbf{Slayer NEO} \cite{MaiorcaICISSP}} & \multicolumn{1}{c|}{Static} & PeePDF+Origami & Structural  & Adaboost \\
\multicolumn{1}{l|}{\textbf{PDFRate}~\cite{Smutz}} & \multicolumn{1}{c|}{Static} & Custom & Structural & Random Forest \\
\multicolumn{1}{l|}{\textbf{PDFRate (updated)}~\cite{DBLP:conf/ndss/SmutzS16}} & \multicolumn{1}{c|}{Static} & Custom & Structural & Classifier Ensemble \\
\bottomrule
\end{tabular}
\end{table*}

\subsection{Pre-processing}

As many other malware detection tools, the first step of PDF malware detectors is to analyze PDF files \emph{statically} and/or \emph{dynamically}.
In the former case, the file is not executed, and information is extracted solely based on static code inspection (typically, through \emph{parsing} the code). In the latter case, suspicious PDF files are dynamically executed through \emph{sandboxing}, in protected virtual environments, and their behavior is monitored.
Dynamic analysis is usually more effective at detecting malicious files, especially when the embedded malicious code has been obfuscated to compromise static analysis. However, dynamic analysis is normally very computationally demanding in terms of both space and time resources, and it may be evaded by other techniques, like a delayed execution of the malicious exploitation code.
In the following, we provide an overview of the tools and libraries that are typically used to extract data from PDF files in current PDF malware detection systems.

\subsubsection{Pre-processing with Third-party Software}

PDF malware detectors based on dynamic analysis normally use sandboxing or code instrumentation (\eg, \texttt{JSand} or \texttt{PhoneyPDF}~\cite{Cova,corona14-aisec}). 

Conversely, detection systems based on static analysis have adopted a variety of solutions over the years. \texttt{Slayer} relies upon \texttt{PDFiD} from PDF files~\cite{Maiorca}. Its updated version (Slayer NEO~\cite{MaiorcaICISSP}), instead, uses \texttt{PeePDF} for a more in-depth analysis of embedded files, multiple versions, and streamed objects, and \texttt{Origami} to perform integrity checks on the file structure and content. These analysis are useful to detect PDF malware hidden with subtle embedding techniques, including anomalous or malformed files. 

Library-based parsing relies on specific PDF libraries that can also be used by open-source PDF readers. The most popular example is \texttt{Poppler}, a comprehensive PDF library that is adopted by the popular open-source reader \texttt{XPDF}. \texttt{PJscan}~\cite{Laskov} and \texttt{Hidost}~\cite{Srndic16} use \texttt{Poppler} to detect PDF files embedding malicious JavaScript code.
Although these libraries correctly implement most of the Adobe PDF specifications, they may be vulnerable to well-crafted malformations of PDF files. 

\subsubsection{Custom Pre-processing}
We refer to pre-processing analyses which do not leverage any third-party PDF-specific tool or library as \emph{custom} pre-processing. Typically, it consists of implementing a static, custom parser to pre-process the input PDF files.
This choice has the advantage of avoiding potential vulnerabilities of existing libraries, \eg, if they do not correctly handle some malformed files.
\texttt{PDFRate} is a good example of a PDF malware detector exploiting a custom parsing mechanism~\cite{Smutz,DBLP:conf/ndss/SmutzS16}.
However, custom parsing itself may introduce other vulnerabilities, if it does not properly follow the Adobe PDF specifications. For example, Adobe Reader completely ignores any object that is not referenced by the \emph{x-ref table} in a PDF file.
Conversely, \texttt{PDFRate} parses those objects. This misbehavior has been exploited in~\cite{Srndic14} to evade \texttt{PDFRate}, through injection of well-crafted objects into PDF malware files. Since these objects are ignored by the reader, they would not compromise the malicious functionality of the embedded exploitation code, while enabling evasion of the detection system. We refer the reader to~\cite{carmony16} for an in-depth evaluation of the vulnerabilities of PDF parsing tools.

\subsection{Feature Extraction}
To classify PDF files as legitimate or malicious using a learning-based algorithm,
a preliminary, required step is to represent each file as a numerical vector of fixed size. This process is usually referred to as \emph{feature extraction}.
 
\subsubsection{JavaScript-based Features}
The vast majority of PDF malware relies on the embedding of malicious JavaScript code.
For this reason, specific features have been exploited to detect evidence of such behavior. The detection approach named \texttt{PJScan}~\cite{Laskov} aims to detect the presence of malicious (obfuscated) JavaScript code by considering occurrences of suspicious API calls like \texttt{eval} or \texttt{replace}, and of string-chaining operators like \texttt{+}, among others.
\texttt{Lux0R}~\cite{corona14-aisec} leverages code instrumentation to detect the presence of API calls in JavaScript code that are specifically used for PDF-related operations. 
\texttt{Wepawet} dynamically executes the embedded JavaScript code using \texttt{JSand}, and then extracts features mostly related to method calls and shellcode memory allocation.

\subsubsection{Structural Features}
Structural features are only related to the characteristics of the \emph{name objects} present in the PDF file.
They do not consider any analysis of the embedded exploitation code.
This has the advantage of being sufficiently general to detect PDF malware embedding different malicious code (\eg, JavaScript or ActionScript). 
However, since the malicious code is not analyzed at all, it is likely that such features can be easily misled by constructing PDF files with similar objects to those typically appearing in legitimate files. PDF malware detectors based on such features include \texttt{Slayer} and \texttt{Slayer NEO}~\cite{Maiorca,MaiorcaICISSP}, \texttt{Hidost}~\cite{Srndic16}, and \texttt{PDFRate}~\cite{Smutz,DBLP:conf/ndss/SmutzS16}.

\subsection{Learning and Classification}

Independently from the chosen feature representation, after feature extraction, each PDF file is represented in terms of a numerical vector $\vct x \in \mathbb R^{\con d}$.
This abstraction enables using any kind of learning algorithm to perform classification of PDF documents, as described in the following.
First, a learning algorithm is \emph{trained} to recognize a set of known examples $\set D \in \{\vct x_{i}, y_{i}\}_{i=1}^{\con n}$, labeled either as legitimate ($y=-1$) or as malicious ($y=+1$). During this process, the parameters of the learning algorithm (if any) are typically set according to some given performance requirements.
After training, the learning algorithm provides a classification function $f(\vct x) \in \{-1,+1\}$ that can be used to classify never-before-seen PDF files as legitimate or malicious.
Clearly, the selection of an appropriate learning algorithm depends on the given data, and on the feature representation. Accordingly, one normally tests different algorithms and retains the one that best fits the given application requirements. The PDF malware detectors mentioned throughout this article adopt different learning algorithms; \eg, \texttt{Wepawet}~\cite{Cova} uses a Bayesian classifier, \texttt{PJScan}~\cite{Laskov} uses Support Vector Machines (SVMs), while several other approaches use classifier ensembles including Random Forests and Adaboost~\cite{Maiorca,MaiorcaICISSP,corona14-aisec,Smutz,Srndic16}, also to improve resilience against some kinds of attack~\cite{DBLP:conf/ndss/SmutzS16}.

\section{Evading Learning-based PDF Malware Detection}
\label{sect:adv-attacks}

Learning-based PDF malware detection has been shown to be effective in detecting malware samples in the wild.
However, it is natural to expect that the level of sophistication of the next generation of attacks will increase again, exploiting vulnerabilities of the architectural components of the detection system that we depicted in Figure~\ref{fig:system-outline}, including the learning algorithm, as envisaged in~\cite{biggio14-tkde,Srndic14}. 
In a typical evasion setting, the attacker's goal is to evade classifier detection by manipulating malware under the constraint that it preserves its intrusive functionality, according to a given level of knowledge of the targeted system. In general, the attacker may know, partially or completely, part of the training data used to learn the classification function, how features are computed from PDF files, and which learning algorithm is used. 

Different attacks against PDF malware detectors have been recently proposed~\cite{Smutz,biggio13-ecml,MaiorcaASIACCS,corona14-aisec,Srndic14,Xu16}.
In terms of the attacker's capability, they only consider the \emph{injection} of different kinds of content into a PDF malware sample.
Removing objects is typically avoided, to keep the functionality of the exploitation code intact. 
In terms of the attacker's knowledge, \emph{mimicry}~\cite{Smutz,corona14-aisec} and \emph{reverse mimicry}~\cite{MaiorcaASIACCS} attacks do not exploit any specific knowledge of the attacked system. In both cases, the content of a benign file is embedded into a malicious PDF, or vice-versa.
In particular, in a \emph{mimicry} attack, the attacker injects benign content (\ie, content extracted from one or more benign PDF files) in a malicious file, to increase the probability of evading detection. 
Conversely, in a \emph{reverse mimicry} attack, the malicious content is injected into a benign file.
More sophisticated attacks, usually referred to as \emph{evasion} attacks, have been proposed against learning-based PDF malware detectors in~\cite{biggio13-ecml,Srndic14}.
These attacks exploit knowledge of the feature set and of the classification function 
to minimize the number of modifications required to evade detection, while maximizing the probability of evasion. 
We refer the reader to \cite{biggio13-ecml,biggio14-tkde,Srndic14,Kantchelian16} for further details on how to implement such attacks.

\subsection{Content Injection in PDF Files}
\label{sect:pdf-embed}

\begin{figure}[t]
	\centering
	\includegraphics[width=0.45\textwidth]{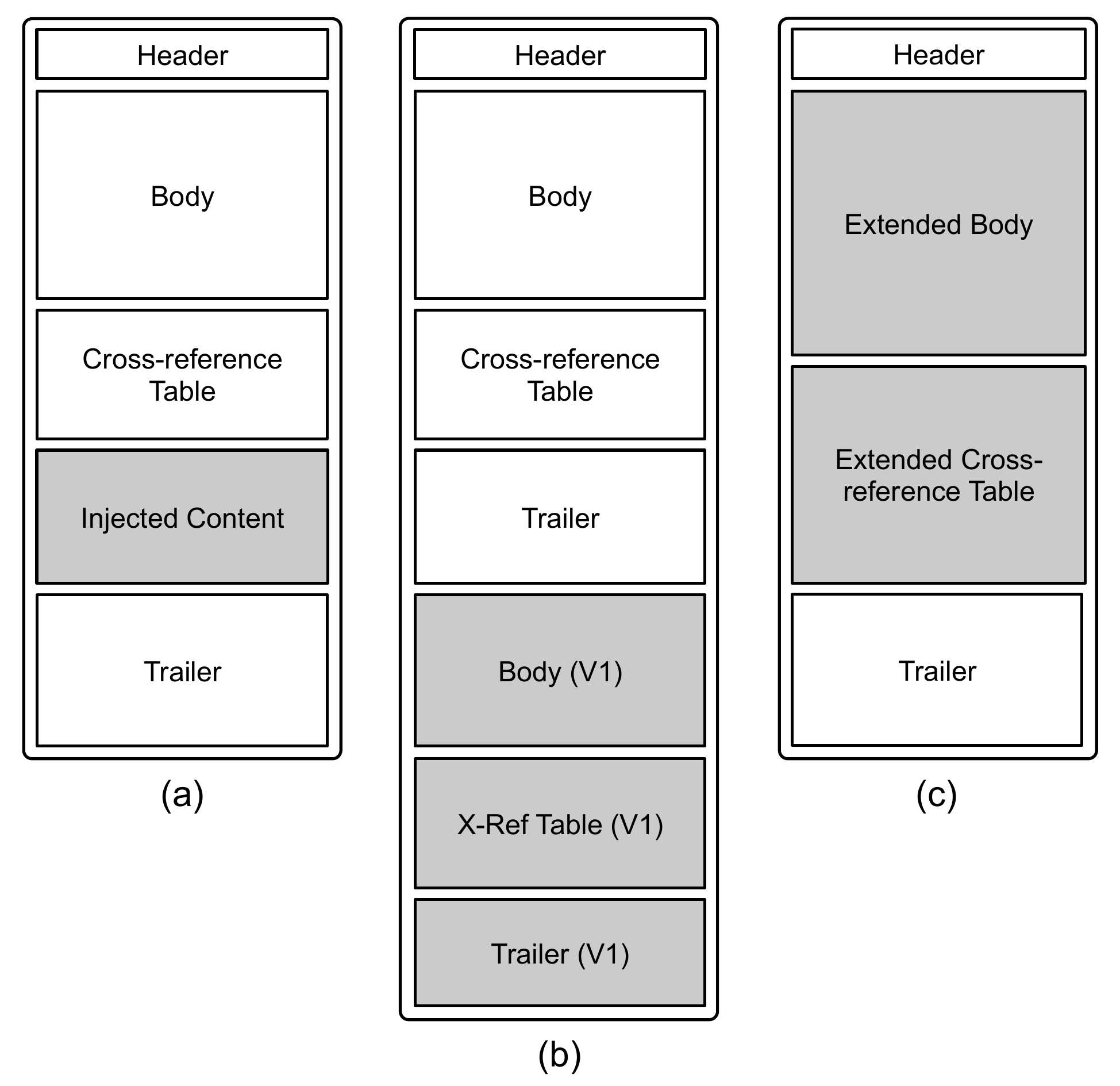}
	\caption{Content injection in PDF files: (a) injecting objects after the x-ref table; 
	(b) using the \emph{versioning} mechanism of the PDF file format; and (c) adding objects to the file body and extending the x-ref table accordingly.}
	\label{sect:pdf-embed:fig:injection}
\end{figure}

Three different techniques can be used to inject content into a PDF file, as conceptually depicted in Figure~\ref{sect:pdf-embed:fig:injection}:
($a$) injecting objects after the x-ref table, as done in~\cite{Srndic14} to evade \texttt{PDFRate};
($b$) using the \emph{versioning} mechanism of the PDF file format, \ie, injecting a new body, x-ref table and trailer, as if the file was directly modified by the user (\eg, by using an external tool);
and ($c$) directly acting on the existing PDF graph, adding new objects to the file body and re-arranging the x-ref table accordingly.

The first strategy is easy to implement, but it can be made ineffective by simply patching the pre-processing module of the PDF malware detector to be consistent with Adobe Reader. In fact, within this strategy the injected content is ignored by Adobe Reader, but not by the pre-processing module of \texttt{PDFRate}. This strategy can clearly be used only in mimicry and evasion attacks, to add benign content to a malicious PDF file.
The other two strategies, instead, can be used to perform reverse mimicry attacks, by injecting malicious code into a benign PDF file. The second strategy is easier to implement, but, clearly, also easier to spot, as it would suffice to correctly extract the additional versions embedded in the file and process them separately.
The third strategy is more complex to implement and to detect, as it seamlessly adds objects in a PDF file yielding a PDF which is essentially indistinguishable from a newly-created one.
It can be implemented using \texttt{Poppler} to manage and re-arrange the x-ref table objects without corrupting the file. Existing objects can also be modified by adding other name objects and rearranging the x-ref table positions accordingly. Notably, it is important to ensure that the embedded content (\ie, the exploitation code) is correctly executed when the merged PDF is opened. This is not an easy task, as it requires injecting additional objects specifically for this purpose. 

\subsection{Empirical Results on Detection Systems}

\begin{figure}[t]
	\hspace{-1.3cm}
	\includegraphics[width=0.55\textwidth]{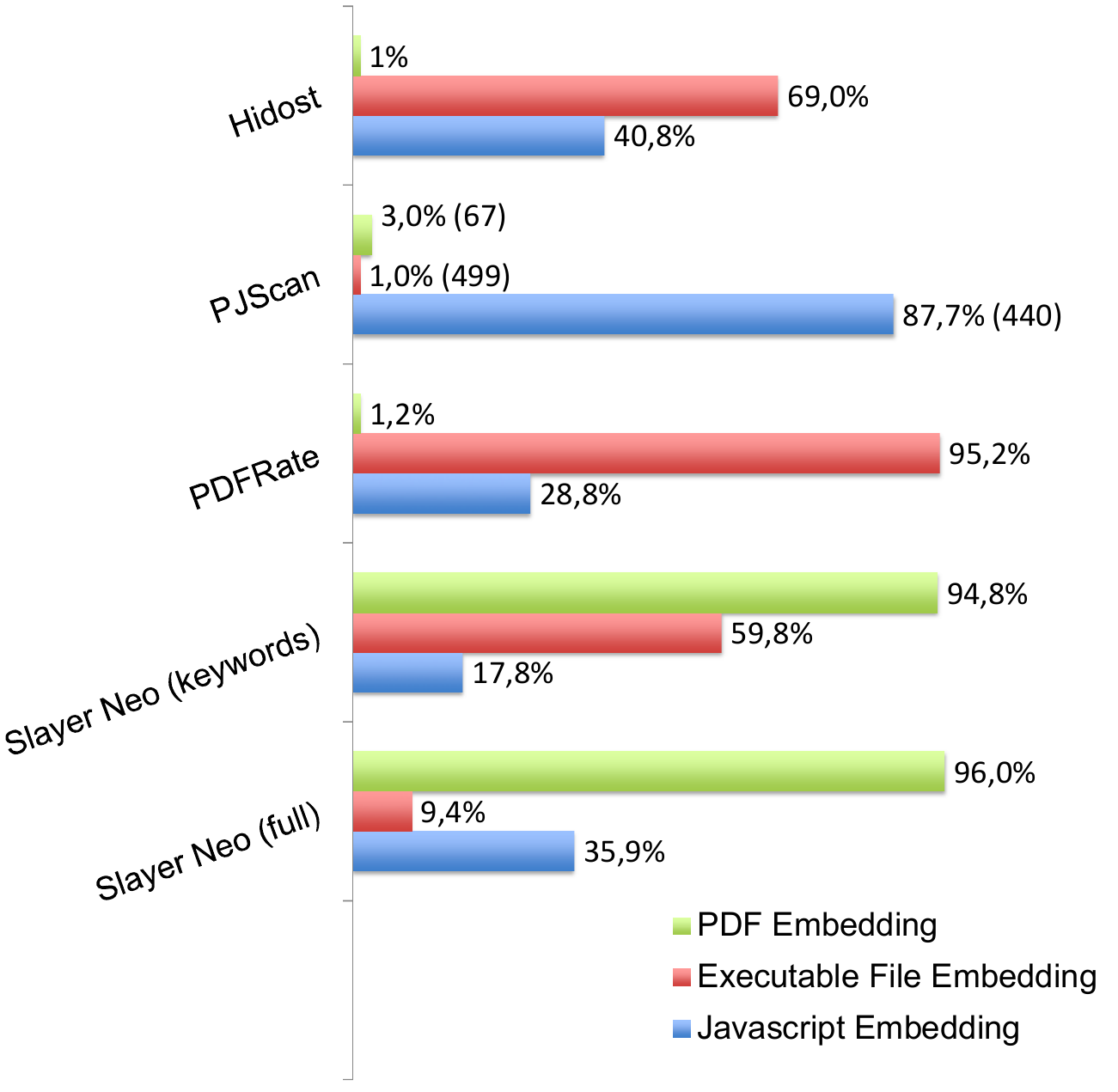}
	\caption{Detection rate of state-of-the-art PDF malware detectors against reverse mimicry attacks embedding different content, using $500$ files per attack. Due to parsing problems, the detection rate of \texttt{PJScan} is estimated on a subset of files, as reported in parentheses.}
	\label{fig:results}
\end{figure}

We report here an empirical evaluation of PDF malware detection tools against reverse mimicry attacks, which only require a limited number of structural changes to the benign source file (with respect to other content-injection attacks).
Content injection in reverse mimicry can be performed with the techniques ($b$) and ($c$) depicted in Figure~\ref{sect:pdf-embed:fig:injection}.

We consider here injection of three different types of content: ($i$) a malicious JavaScript exploitation routine, ($ii$) a malicious PDF file, and ($iii$) a malicious executable (\ie, the Zeus trojan payload).

JavaScript embedding was performed by injecting the same malicious code in different benign files, for two reasons: (a) we wanted to verify whether different PDF file structures could influence the detection of the same malware; (b) the current version of this embedding procedure only supports malicious codes contained in one single object.
More advanced attacks usually involve spreading JavaScript codes in multiple objects. It would have been therefore unfeasible to use different malicious codes.
In the PDF embedding attack, we injected one random malicious file (gathered from \texttt{VirusTotal}\footnote{\url{http://www.virustotal.com}}) into each benign file. We wrote efficient injection routines for both JavaScript and PDF embedding attacks by employing \texttt{Poppler}. 
EXE embedding was performed using Metasploit to automatically inject a malicious payload in each benign file.  

Each of the aforementioned malicious contents was hidden into $500$ different benign PDF files (gathered from the \texttt{Yahoo} search engine), yielding a complete dataset of $1,500$ reverse mimicry attacks, which are publicly available.\footnote{\url{https://pralab.diee.unica.it/en/pdf-reverse-mimicry/}} \texttt{PJScan}, \texttt{Hidost} and \texttt{Slayer NEO} were all trained with a dataset composed by more than $20000$ malicious and benign files, respectively collected from \texttt{VirusTotal} and the \texttt{Yahoo} search engine. For the sake of a fair comparison, we also trained \texttt{Slayer NEO} and \texttt{Hidost} with the same classification algorithm (\texttt{Adaboost}). Note also that \texttt{Slayer NEO} was used by employing both the algorithm described in \cite{Maiorca} (keywords) and the one described in \cite{MaiorcaICISSP} (keywords and content-based features).

The results are reported in Figure~\ref{fig:results}. 
\texttt{PDFRate}, \texttt{Slayer Neo}, and \texttt{Hidost} are especially effective at detecting EXE embedding attacks, as they introduce specific keywords. However, they struggle at detecting JavaScript-based attacks, as PDF structures that are apparently malicious in terms of keywords can simply contain benign code. Content-based systems are more effective at detecting embeddings of JavaScript code, but they might fail under specific circumstances. \texttt{PJScan}, in particular, suffers from the presence of multiple embedded JavaScript codes, which may happen when embedding a malicious script into a benign file that already contains JavaScript codes.

With respect to PDF embedding attacks, \texttt{Slayer NEO} is the only effective system that can detect them, as it automates the analysis of embedded files. In particular, as the system extracts and analyzes embedded files separately from their benign containers, its detection capabilities are not influenced by the presence of benign features.   

In conclusion, we can state that there is no unique solution for detecting all the attacks. Each tool should be considered to perform a thorough digital investigation. 


\section{Summary and Open Problems}

In the last decades, fueled by a flourishing underground economy, malware has grown exponentially, not only in terms of the mere number of variants and families, but also in terms of sophistication, mainly to evade current detection approaches. 
In this article, we have discussed how the PDF file format can be exploited by attackers to convey malware, leveraging the possibility of embedding different kinds of content, and, accordingly, of exploiting different, potential vulnerabilities. We have provided practical examples of known malware and zero-day exploits, and discussed current detection systems based on machine learning.
We believe that such systems can be extremely helpful for a forensic analyst to understand the suspiciousness of a PDF file, and the potential root causes behind infection. 
Envisaging the next step of the arms race between malware and system developers, we have then discussed the security properties of learning algorithms against well-crafted evasion attempts, reporting also some empirical results. This is another important aspect besides improving the security of other system components like pre-processing and parsing, since machine-learning algorithms exhibit intrinsic vulnerabilities that will be, sooner or later, exploited by skilled and economically-motivated attackers. 
In a security-by-design perspective, being proactive demands for the development of \emph{adversarial learning machines}, \ie, learning algorithms that explicitly account for the presence of malicious input data manipulations and provide improved security guarantees~\cite{biggio14-tkde,Srndic14,Kantchelian16}.
This may definitely be one of the most relevant research challenges in the coming years.



\begin{IEEEbiography}
[{\includegraphics[width=1in,height =1.25in,clip,keepaspectratio]{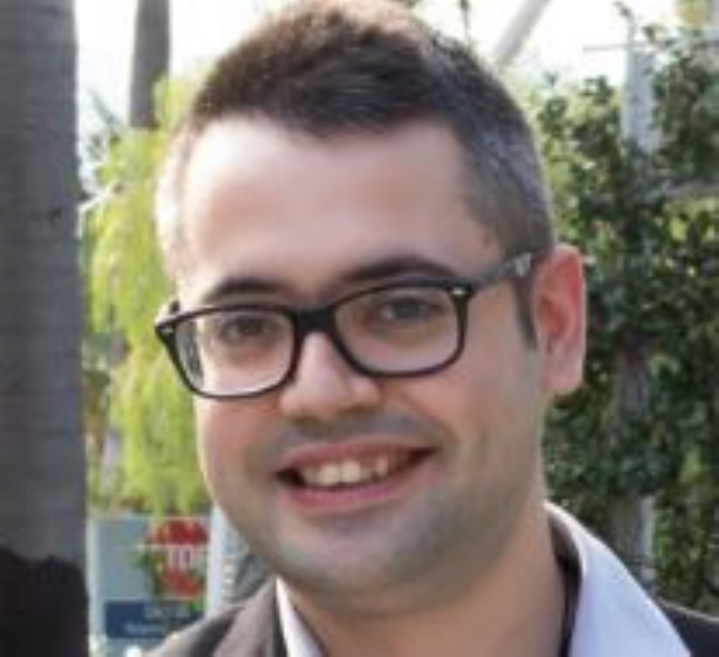}}]{Davide Maiorca (M'16)} received from the University of Cagliari (Italy) the M.Sc. degree (Hons.) in Electronic Engineering in 2012 and the Ph.D. in Electronic Engineering and Computer Science in 2016. In 2013, he visited the Systems Security group at Ruhr-Universit\"at Bochum, guided by Prof. Dr. Thorsten Holz, and worked on advanced obfuscation of Android malware. His current research interests include adversarial machine learning, malware in documents and Flash applications, Android malware and mobile fingerprinting. He has been a member of the 2016 IEEE Security \& Privacy Student Program Committee.
\end{IEEEbiography}

\begin{IEEEbiography}
[{\includegraphics[width=1in,height =1.25in,clip,keepaspectratio]{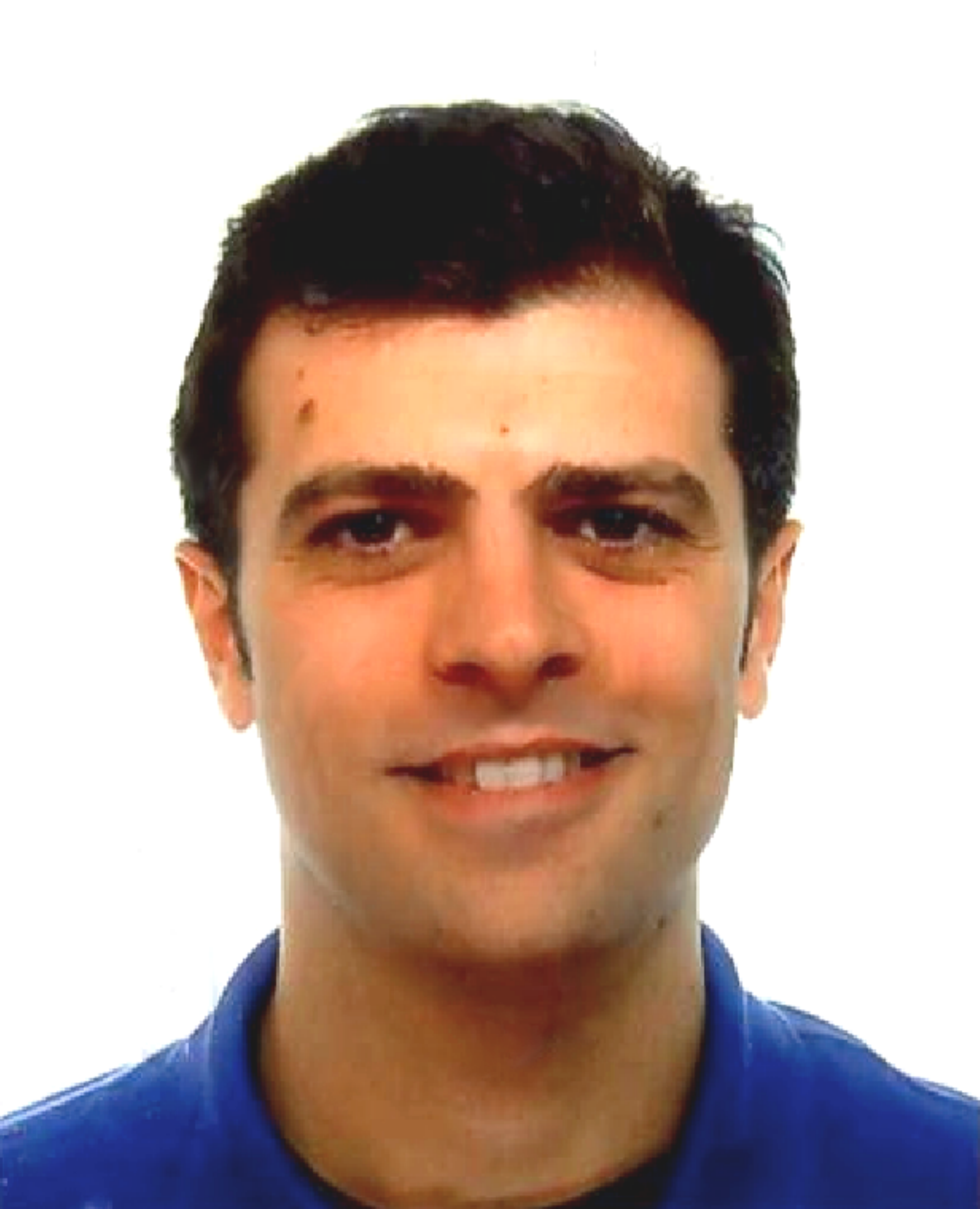}}]{Battista Biggio (SM'17)} received the M.Sc. degree (Hons.) in Electronic Engineering and the Ph.D. degree in Electronic Engineering and Computer Science from the University of Cagliari, Italy, in 2006 and 2010. Since 2007, he has been with the Department of Electrical and Electronic Engineering, University of Cagliari, where he is currently an Assistant Professor. In 2011, he visited the University of T\"ubingen, Germany, and worked on the security of machine learning to training data poisoning. His research interests include secure machine learning, multiple classifier systems, kernel methods, biometrics and computer security. Dr. Biggio serves as a reviewer for several international conferences and journals. He is a senior member of the IEEE and a member of the IAPR.
\end{IEEEbiography}
\end{document}